\documentclass[11pt]{article}

\usepackage{amsmath}
\usepackage{amsfonts}
\usepackage{amssymb}
\usepackage{amscd}
\usepackage{latexsym}
\usepackage{mathrsfs}
\usepackage{amsthm}

\newtheorem{Theorem}{Theorem}
\newtheorem{Proposition}[Theorem]{Proposition}
\newtheorem{Lemma}[Theorem]{Lemma}
\newtheorem{Remark}[Theorem]{Remark}

\def\C{\mathbb{C}}

\def\H{\mathcal{H}}
\def\G{\mathcal{G}}
\def\R{\mathbb{R}}

\def\S{\mathbb{S}}
\def\T{\mathbb{T}}
\def\Z{\mathbb{Z}}
\def\d{\mathrm{d}}
\def\F21{{}_2F_1}

\def\Pv{\mathrm{Pv}}
\def\FF{\mathcal{F}}
\def\Opm{\Omega_{\pm,m}^{A\!B}}

\def\li{\mathrm{l.i.m.}}

\def\AB{A\!B}
\def\Tm{\mathscr{T}}
\def\Sm{\mathscr{S}}
\def\2{\mathfrak{int}}
\def\k{\textsc k}

\begin{document}

\title{New formulae for the Aharonov-Bohm wave operators}

\author{S. Richard$\footnote{On leave from Universit\'e de Lyon, Universit\'e Lyon 1, CNRS, UMR 5208, Institut Camille Jordan, 43 blvd du 11
novembre 1918, F-69622 Villeurbanne-Cedex, France.}\ $}
\date{\small}
\maketitle \vspace{-1cm}

\begin{quote}
\emph{
Department of Pure Mathematics and Mathematical Statistics,
Centre for Mathematical Sciences, University of Cambridge,
Cambridge, CB3 0WB, United Kingdom.
\emph{E-mail:} sr510@cam.ac.uk}
\end{quote}

\begin{abstract}
It is proved that the wave operators corresponding to Schr\"odinger operators with Aha\-ro\-nov-Bohm type magnetic fields can be rewritten in terms of explicit functions of the generator of dilations and of the Laplacian.
\end{abstract}

\section{Introduction}

In some recent works on scattering theory \cite{KR1,KR2,KR3,KR5}, it was conjectured and then proved that, modulo a compact term, the wave operators for Schr\"odinger systems can be rewritten as a product of a function of the dilation operator $A$ and a function of the Laplacian $-\Delta$. Furthermore, the functions of the dilation operator are rather insensitive to a particular choice of the perturbed operator and depend mainly on the free system and on the space dimension.

In this paper, we obtain a similar result for the five-parameter family of Hamiltonians describing the non-relativistic Aharonov-Bohm systems \cite{AT,DS}. More precisely, we first show that the wave operators for the original Aharonov-Bohm Hamiltonian \cite{AB,Rui} can be rewritten as explicit functions of $A$ only. For the wave operators corresponding to other self-adjoint extensions, we prove that the additional terms are given by the product of a function of $A$ and a function of $-\Delta$. Let us already stress that the functions of the dilation operator depend on the flux of the magnetic field, but not on the other parameters of the boundary condition at $0\in \R^2$.

These new formulae might serve for various further investigations on scattering theory for systems with less singular magnetic fields. In particular, it would interesting to study the structure of the wave operators for Schr\"odinger operators with magnetic fields supported in small sets, see for example \cite{EIO,T1,T2}. These new expressions also lead to a topological approach of Levinson's theorem. In this respect, we mention two papers related to Levinson's for the original Aharonov-Bohm operator \cite{L,SM}. We intend to address both subjects in forthcoming publications.

The structure of this paper is the following: We first recall the constructions of the five-parameter family of self-adjoint operators, mainly borrowed from \cite{AT}. After a technical interlude on the Fourier transform and on the generator of dilations, we show in Theorem \ref{thm1} that the wave operators for the original Aharonov-Bohm system can be rewritten as functions of $A$ only. We then extend the analysis to the wave operators for  arbitrary self-adjoint extensions, and propose new formulae for them in Theorem \ref{thm2}.

\noindent
{\bf Acknowledgments:} This work is supported by the Swiss National Science Foundation. Its author also thanks S. Nakamura for a two weeks invitation to Japan where part of the present work was completed. This stay was made possible thanks to a grant from the Japan Society for the Promotion of Science.

\section{The family of self-adjoint extensions}\label{secex}

In this section, we recall the construction of \cite{AT} for the family of self-adjoint extensions corresponding to a  Schr\"odinger operator with a singular magnetic field located at the origin. We also refer to \cite{DS} for a similar construction, and to \cite{T1,T2} for more details.

Let us set $\H$ for the Hilbert space $L^2(\R^2)$ and denote its norm by $\|\cdot \|$. For any $\alpha \in (0,1)$, we define $A_\alpha: \R^2\setminus\{0\} \to \R^2$ by
\begin{equation*}
A_\alpha(x_1,x_2)= -\alpha\left(\frac{-x_2}{x_1^2+x_2^2},
\frac{x_1}{x_1^2+x_2^2}\right)\
\end{equation*}
and consider the operator
\begin{equation*}
H_\alpha:=(-i\nabla -A_\alpha)^2, \qquad
D(H_\alpha)=C_c^\infty\big(\R^2\setminus\{0\}\big)\ ,
\end{equation*}
where $C_c^\infty(\Xi)$ denotes the set of smooth functions on $\Xi$ with compact support.
The closure $\overline{H}_\alpha$ of this operator in $\H$ is symmetric and has deficiency indices $(2,2)$. The deficiency subspace $\Sigma_+:=\ker (\overline{H}_\alpha^*- i)$ is spanned by the functions (in polar coordinates)
\begin{equation*}
\psi_+^0(r):=c_+^0 \;\!K_\alpha(e^{-i\pi/4}r)\;\!\hbox{$\frac{1}{\sqrt{2\pi}}$}
\quad \hbox{ and } \quad
\psi_+^{-1}(r,\theta):=c_+^{-1}\;\!K_{1-\alpha}(e^{-i\pi/4}r)\;\!\hbox{$\frac{e^{-i\theta}}{\sqrt{2\pi}}$}\ ,
\end{equation*}
and the deficiency subspace $\Sigma_-:=\ker (\overline{H}_\alpha^*+ i)$ is spanned by the functions
\begin{equation*}
\psi_-^0(r):=c_-^0\;\!e^{i\pi\alpha/2}\;\!K_\alpha(e^{i\pi/4}r)\;\!\hbox{$\frac{1}{\sqrt{2\pi}}$}
\quad \hbox{ and } \quad
\psi_-^{-1}(r,\theta):=c_-^{-1}\;\!e^{i\pi (1-\alpha)/2}\;\!K_{1-\alpha}(e^{i\pi/4}r)\;\!\hbox{$\frac{e^{-i\theta}}{\sqrt{2\pi}}$}\ .
\end{equation*}
Here, $K_\mu$ is the modified Bessel function of the second kind and of order $\mu$, and the real constants $c_\pm^0$ and $c_\pm^{-1}$ are chosen such that $\|\psi_\pm^0\|=\|\psi_\pm^{-1}\|=1$.

By the standard theory of Krein, all self-adjoint extensions of $\overline{H}_\alpha$ are parameterized by the set of unitary maps from one deficiency subspace to the other one. Therefore, for $\eta\in \R$ and $a,b \in \C$ satisfying $|a|^2+|b|^2=1$, let us set
\begin{equation*}
U = U(\eta,a,b)=e^{i\eta}\left(
\begin{array}{cc}
a & -\overline{b}\\
b & \overline{a}
\end{array}\right)
\end{equation*}
for a general unitary map from $\Sigma_+$ to $\Sigma_-$. These spaces are endowed with their respective bases $\{\psi_+^0, \psi_+^{-1}\}$ and $\{\psi_-^0, \psi_-^{-1}\}$.
Then, for any such $U$, there exists a self-adjoint extension $H_\alpha^U$ of $\overline{H}_\alpha$ defined by
\begin{equation*}
D(H_\alpha^U)= \Big\{f \in \H \mid f = g + \psi_+ + U\psi_+ \hbox{ with } g \in D(\overline{H}_\alpha), \psi_+ \in \Sigma_+\Big\}
\end{equation*}
and
\begin{equation*}
H_\alpha^U f = \overline{H}_\alpha g +i\psi_+ - iU\psi_+\ .
\end{equation*}
In particular, the special choice $U(0,-1,0)$ coincides with the
original operator $H_\alpha^{\AB}$ introduced by Aharonov and Bohm and thoroughly studied in \cite{Rui}.

The generalized eigenfunctions of these operators have been calculated in \cite{AT}, and we shall come back to them subsequently. Some useful tools and notations have first to be introduced.

\section{Fourier transform and dilation operator}

In this section we briefly recall the necessary background on the Fourier transform and the dilation operator.

Let us first decompose the Hilbert space $\H$ with respect to polar coordinates: For any $m \in \Z$, let $\phi_m$ be the complex function defined by
$[0,2\pi)\ni \theta \mapsto \phi_m(\theta):= \hbox{$\frac{e^{im\theta}}{\sqrt{2\pi}}$}$. Then, by taking the completeness of the family $\{\phi_m\}_{m \in \Z}$ in $L^2(\S^1)$ into account, one has the canonical decomposition
\begin{equation*}
\H = \bigoplus_{m \in \Z} \H_m \ ,
\end{equation*}
where $\H_m=\big\{f \in \H \mid f(r,\theta)=g(r)\phi_m(\theta) \hbox{ a.e.~for some }g \in \H_r\big\}$, $\H_r:=L^2(\R_+, r\;\!\d r)$ and $\R_+$ is the open interval $(0,\infty)$.

Let $\FF$ be the usual Fourier transform, explicitly given on any $f \in \H$ and $k \in \R^2$ by
\begin{equation*}
[\FF f](k)= \frac{1}{2\pi}\;\!\li \int_{\R^2}f(x)\;\!e^{-ix\cdot k}\;\!\d x
\end{equation*}
where $\li$~denotes the convergence in the mean.
Its inverse is denoted by $\FF^*$. Since the Fourier transform maps the subspace $\H_m$ of $\H$ onto itself, we naturally set $\FF_m: \H_r \to \H_r$ by the relation $\FF(g\phi_m) = \FF_m(g)\phi_m$ for any $g \in \H_r$. More explicitly, the application $\FF_m$ is the unitary map from $\H_r$ to $\H_r$ given on any $g \in \H_r$ and $\kappa \in \R_+$ by
\begin{equation*}
[\FF_m g](\kappa)= (-i)^{|m|}\;\!\li\int_{\R_+} r\;\!J_{|m|}(r\;\!\kappa)\;\!g(r)\;\!\d r\ ,
\end{equation*}
where $J_{|m|}$ denotes the Bessel function of the first kind and of order $|m|$. The inverse Fourier transform $\FF_m^*$ is given by the same formula, with $(-i)^{|m|}$ replaced by $i^{|m|}$.

Let us now consider the unitary dilation group $\{U_\tau\}_{\tau \in \R}$ defined on any $f \in \H$ and $x \in \R^2$ by
\begin{equation*}
[U_\tau f](x) = e^\tau f(e^\tau x)\ .
\end{equation*}
Its self-adjoint generator $A$ is formally given by $\hbox{$\frac{1}{2}$}(X\cdot (-i\nabla) + (-i\nabla)\cdot X)$, where $X$ is the position operator and $-i\nabla$ is its conjugate operator. All these operators are essentially self-adjoint on the Schwartz space on $\R^2$.

It is easily observed that the formal equality $\FF\;\!A\;\!\FF^*=-A$ holds. More precisely, for any essentially bounded function $\varphi$ on $\R$, one has $\FF\varphi(A)\FF^* = \varphi(-A)$. Furthermore, since $A$ acts only on the radial coordinate, the operator $\varphi(A)$ leaves each $\H_m$ invariant.
So, for any $m \in \Z$, let $\varphi_m$ be an essentially bounded function on $\R$. Assume furthermore that the family $\{\varphi_m\}_{m \in \Z}$ is bounded. Then the operator $\varphi(A):\H \to \H$ defined on $\H_m$ by $\varphi_m(A)$ is a bounded operator.

Let us finally recall a general formula about the Mellin transform.
\begin{Lemma}\label{Jensen}
Let $\varphi$ be an essentially bounded function on $\R$ such that its inverse Fourier transform is a distribution on $\R$. Then, for any $f \in C^\infty_c\big(\R^2\setminus\{0\}\big)$ one has
\begin{equation*}
[\varphi(A)f](r,\theta) =
\big(\hbox{$\frac{1}{2\pi}$}\big)^{1/2} \int_0^\infty\check{\varphi}
\big(-\ln(\hbox{$\frac{s}{r}$})\big)\;\!f(s,\theta)\;\! \hbox{$\frac{\d s}{r}$}\ ,
\end{equation*}
where the r.h.s.~has to be understood in the sense of distributions.
\end{Lemma}

\begin{proof}
The proof is a simple application for $n=2$ of the general formulae developed in \cite[p.~439]{Jen}.
Let us however mention that the convention of this reference on the minus sign for the operator $A$ in its spectral representation
has not been adopted.
\end{proof}

As already mentioned $\varphi(A)$ leaves $\H_m$ invariant. And more precisely, if $f=g\phi_m$ for some $g \in C^\infty_c(\R_+)$, then $\varphi(A)g\phi_m = [\varphi(A)g]\phi_m$ with
\begin{equation}\label{formuleJen}
[\varphi(A)g](r) =
\big(\hbox{$\frac{1}{2\pi}$}\big)^{1/2} \int_0^\infty\check{\varphi}
\big(-\ln(\hbox{$\frac{s}{r}$})\big)\;\!g(s)\;\! \hbox{$\frac{\d s}{r}$}\ ,
\end{equation}
where the r.h.s.~has again to be understood in the sense of distributions.

\section{The original Aharonov-Bohm operator}\label{secABop}

Let us now come back to the original Aharonov-Bohm operator $H_\alpha^{\AB}$. We shall recall some formulae gathered in the paper \cite{Rui}. For shortness, the index $\alpha$ will be omitted in certain expressions. Since the operator $H_\alpha^{\AB}$ leaves each subspace $\H_m$ invariant, it gives rise to a sequence of channel operators $H_{\alpha,m}^{\AB}$ acting on $\H_m$.
The usual operator $-\Delta$ admitting a similar decomposition, the wave operators
\begin{equation*}
\Omega_\pm^{\AB} :=s-\lim_{t\to\pm \infty}e^{iH_\alpha^{\AB} t}\;\!e^{-i(-\Delta)t}\ .
\end{equation*}
can be defined in each channel, {\it i.e.}~separately for each $m \in \Z$. Let us immediately observe that the angular part does not play any role for defining such operators. Therefore, we shall omit it as long as it does not lead to any confusion, and consider the channel wave operators
$\Opm$ from $\H_r$ to $\H_r$. It is proved in \cite[Thm.~A1]{Rui}
that these operators exist and are isometric maps from $\H_r$ onto $\H_r$. Furthermore, they are given for any $g \in \H_r$ and $r \in \R_+$ by
\begin{equation}\label{formuleRui}
[\Opm\;\! g](r)= i^{|m|}\;\li\int_{\R_+} \kappa\;\!
J_{|m+\alpha|}(\kappa\;\!r)\;\!e^{\mp i\delta_m^\alpha}\;\![\FF_m g](\kappa)\;\!\d
\kappa\ ,
\end{equation}
where
\begin{equation*}
\delta_m^\alpha = \hbox{$\frac{1}{2}$}\pi\big(|m|-|m+\alpha|\big)
=\left\{\begin{array}{rl}
-\hbox{$\frac{1}{2}$}\pi\alpha & \hbox{if }\ m\geq 0 \\
\hbox{$\frac{1}{2}$}\pi\alpha & \hbox{if }\ m< 0
\end{array}\right.\ .
\end{equation*}

Since the wave operators admit a decomposition into channel wave operators, so does the scattering operator. The channel scattering
operator $S_m^{\AB}:=(\Omega_{+,m}^{\AB})^*\;\!\Omega_{-,m}^{\AB}: \H_r\to \H_r$ is simply given by \cite[eq.~4.6]{Rui}~:
\begin{equation*}
S_m^{\AB} = e^{2i\delta_m^\alpha}\ .
\end{equation*}

Let us now concentrate on the channel wave operators. Since $C^\infty_c(\R_+)$ is contained in $\H_r$, one has for any $g \in C^\infty_c(\R_+)$ and $r \in \R_+$:
\begin{eqnarray}\label{fgeneral}
\nonumber [\Opm g](r) &=& s-\lim_{N\to \infty}i^{|m|}\int_0^N \kappa\;\!J_{|m+\alpha|}(\kappa\;\!r)\;\!
e^{\mp i\delta_m^\alpha}[\FF_m g](\kappa)\;\!\d \kappa\\
\nonumber &=& s-\lim_{N\to \infty}e^{\mp i\delta_m^\alpha}\int_0^N \kappa \;\!J_{|m+\alpha|}(\kappa\;\!r)
\Big[\int_0^\infty s\;\! J_{|m|}(s\;\!\kappa)\;\!g(s)\;\!\d s\Big]\d \kappa \\
\nonumber &=& s-\lim_{N\to \infty}e^{\mp i\delta_m^\alpha} \int_0^\infty s\;\!g(s)
\Big[\int_0^N \kappa\;\! J_{|m|}(s\;\!\kappa)\;\!J_{|m+\alpha|}(\kappa\;\!r)\;\!\d \kappa\Big] \d s \\
\nonumber &=& s-\lim_{N\to \infty}e^{\mp i\delta_m^\alpha}\int_0^\infty \hbox{$\frac{s}{r}$}\;\!g(s)
\Big[\int_0^{Nr} \kappa\;\! J_{|m|}(\hbox{$\frac{s}{r}$}\;\!\kappa)\;\!J_{|m+\alpha|}(\kappa)\;\!\d \kappa\Big] \hbox{$\frac{\d s}{r}$} \\
&=& e^{\mp i\delta_m^\alpha} \int_0^\infty \hbox{$\frac{s}{r}$} \Big[
\int_0^\infty \kappa\;\! J_{|m|}(\hbox{$\frac{s}{r}$}\;\!\kappa)
\;\!J_{|m+\alpha|}(\kappa)\;\!\d \kappa\Big]g(s)\;\!  \hbox{$\frac{\d s}{r}$}\ ,
\end{eqnarray}
where the last term has to be understood in the sense of distributions on $\R_+$.

Our interest in rewriting the channel wave operators in this form is twofold. Firstly, by comparing \eqref{fgeneral} with \eqref{formuleJen}, one observes that the channel wave operator $\Opm$ is equal, at least on a dense set in $\H_r$, to $\varphi_m^\pm(A)$ for a function $\varphi_m^\pm$ whose inverse Fourier transform satisfies for $y\in\R$:
\begin{equation*}
\check{\varphi}_m^\pm (y) =
\sqrt{2\pi}\;\!e^{\mp i\delta_m^\alpha}\;\!e^{-y}\;\!\Big[
\int_0^\infty \kappa\;\! J_{|m|}(e^{-y}\;\!\kappa)
\;\!J_{|m+\alpha|}(\kappa)\;\!\d \kappa\Big]\ .
\end{equation*}
Secondly, the distribution between brackets has been explicitly computed in \cite[Prop.~2]{KR_WS}. We recall here the general result (the notation $\delta$ is used for the Dirac measure centered at $0$ and $\Pv$ denotes the principal value integral).

\begin{Proposition}\label{JmuJnu}
For any $\mu,\nu \in \R$ satisfying $\nu+2>|\mu|$ and $\mu+2>|\nu|$, and $s \in \R_+$
one has
\begin{eqnarray}\label{horreur2}
&\int_0^\infty \kappa \;\!J_\mu (s\kappa) \;\!J_\nu(\kappa)\;\! \d \kappa\ = \cos(\pi(\nu -\mu)/2)\;\!\delta(s-1)
 + \hbox{$\frac{2}{\pi}$} \;\!\sin(\pi(\nu -\mu)/2) \;\!
s^{-1} \;\!\Pv\;\!\big(\hbox{$\frac{1}{\frac{1}{s}-s}$}\big)\;\! & \\
\nonumber &+ \left\{
\begin{array}{ll}
\hbox{$\frac{2}{\pi}$} \;\!\sin(\pi(\nu -\mu)/2) \;\!
\hbox{$\frac{s^{-1}}{\frac{1}{s}-s}$}\;\!\Big[
s^\mu\hbox{$\frac{\Gamma(\frac{\mu+\nu}{2}+1)\;\!\Gamma(\frac{\mu-\nu}{2}+1)}{\Gamma(\mu+1)}$}\;\!
\F21\big(\hbox{$\frac{\mu+\nu}{2}$},\hbox{$\frac{\mu-\nu}{2}$};\mu+1;s^2\big)-1\Big]
&\mbox{ if } s< 1,\\
\hbox{$\frac{2}{\pi}$} \;\!\sin(\pi(\nu -\mu)/2) \;\!
\hbox{$\frac{s^{-1}}{\frac{1}{s}-s}$}\;\!\Big[
s^{-\nu}\hbox{$\frac{\Gamma(\frac{\nu+\mu}{2}+1)\;\!\Gamma(\frac{\nu-\mu}{2}+1)}{\Gamma(\nu+1)}$}\;\!
\F21\big(\hbox{$\frac{\nu+\mu}{2}$},\hbox{$\frac{\nu-\mu}{2}$};\nu+1;s^{-2}\big)-1\Big]
&\mbox{ if }s> 1,
\end{array}
\right.&
\end{eqnarray}
as an equality between two distributions on $\R_+$. The last term belongs to $L^1_{\rm loc}(\R_+)$.
\end{Proposition}

Thus, let us define the following distributions for $y \in \R$:
\begin{eqnarray*}
\check{\varphi}_{m,1}^\pm(y) &=& \sqrt{2\pi}\;\!e^{\mp i\delta_m^\alpha}
\cos(\delta_m^\alpha)\;\!\delta(y)\ , \\
\check{\varphi}_{m,2}^\pm(y) &=& -\sqrt{\hbox{$\frac{2}{\pi}$}}\;\! e^{\mp i\delta_m^\alpha}\;\!\sin(\delta_m^\alpha) \;\!
\Pv\;\!\big(\hbox{$\frac{1}{\sinh(y)}$}\big)\ .
\end{eqnarray*}
For $y < 0$ let us also set
\begin{equation*}
\check{\varphi}_{m,3}^\pm(y) =-\sqrt{\hbox{$\frac{2}{\pi}$}}\;\!e^{\mp i\delta_m^\alpha}\;\!\sin(\delta_m^\alpha) \;\!
\hbox{$\frac{1}{\sinh(y)}$}\;\!\Big[
e^{\nu y}\hbox{$\frac{\Gamma(\frac{\nu+\mu}{2}+1)\;\!\Gamma
(\frac{\nu-\mu}{2}+1)}{\Gamma(\nu+1)}$}\;\!
\F21\big(\hbox{$\frac{\nu+\mu}{2}$},\hbox{$\frac{\nu-\mu}{2}$}
;\nu+1;e^{2y}\big)-1\Big]\ ,
\end{equation*}
and for $y > 0$
\begin{equation*}
\check{\varphi}_{m,3}^\pm(y) = -\sqrt{\hbox{$\frac{2}{\pi}$}}\;\!e^{\mp i\delta_m^\alpha} \;\!\sin(\delta_m^\alpha) \;\!
\hbox{$\frac{1}{\sinh(y)}$}\;\!\Big[
e^{-\mu y}\hbox{$\frac{\Gamma(\frac{\mu+\nu}{2}+1)
\;\!\Gamma(\frac{\mu-\nu}{2}+1)}{\Gamma(\mu+1)}$}\;\!
\F21\big(\hbox{$\frac{\mu+\nu}{2}$},\hbox{$\frac{\mu-\nu}{2}$}
;\mu+1;e^{-2y}\big)-1\Big]\ ,
\end{equation*}
where the notation $\mu = |m|$ and $\nu = |m+\alpha|$ has been used for shortness. The sum of these distributions is clearly equal to $\check{\varphi}_m^\pm$. These distributions are the inverse Fourier transforms of continuous functions, as proved in the next lemma. We use the notation $\T$ for the set of complex numbers of modulus $1$.

\begin{Lemma}\label{philibre}
One has:
\begin{enumerate}
\item $\varphi_{m,1}^\pm= e^{\mp i\delta_m^\alpha}\cos(\delta_m^\alpha)$,
\item $\varphi_{m,2}^\pm = i\;\!e^{\mp i\delta_m^\alpha}\;\!\sin(\delta_m^\alpha)
\tanh\big(\hbox{$\frac{\pi}{2}\cdot$}\big)$,
\item $\varphi_{m,3}^\pm \in C_0(\R)$ with $\sup_{y \in \R}|\varphi_{m,3}^\pm(y)|\leq 2$ independently of $m \in \Z$.
\end{enumerate}
In particular, one has $\varphi_m^\pm:=\varphi_{m,1}^\pm+\varphi_{m,2}^\pm+\varphi_{m,3}^\pm \in C\big([-\infty,+\infty],\T\big)$, with $\varphi_m^\pm(\pm\infty)= 1$ and $\varphi_m^\pm(\mp\infty) = e^{\mp 2i\delta_m^\alpha}$.
\end{Lemma}

\begin{proof}
The Fourier transform of $\check{\varphi}_{m,1}^\pm$ and $\check{\varphi}_{m,2}^\pm$ are well known. For $\check{\varphi}_{m,3}^\pm$, let us first recall that the two hypergeometric functions appearing in its  definition are bounded functions for $y<0$ and $y>0$, respectively. Thus, the function $y\to \check{\varphi}_{m,3}^\pm(y)$ goes exponentially rapidly to $0$ as $|y|\to \infty$. Finally, it follows from the $L^1_{\rm loc}$-property mentioned in the above proposition that $\check{\varphi}_{m,3}^\pm$ is also locally $L^1$ in a neighbourhood of $y=0$. Altogether one has obtained that
$\check{\varphi}_{m,3}^\pm$ belongs to $L^1(\R)$, and thus its Fourier transform belongs to $C_0(\R)$.

The $L^\infty$-norm of $\varphi_{m,3}^\pm$ and the remaining statements follow from the unitarity of the channel wave operators and some straightforward computations.
\end{proof}

By the density of $C_c^\infty(\R_+)$ in $\H_r$, one has thus obtained:
\begin{Theorem}\label{thm1}
For each $m \in \Z$, one has
\begin{equation*}
\Opm = \varphi_m^\pm(A)\ ,
\end{equation*}
with $\varphi_m^\pm \in C\big([-\infty,+\infty],\T\big)$. These functions are explicitly defined in Lemma \ref{philibre} and above.
\end{Theorem}

\section{Stationary scattering theory}

In this section, we shall be concerned with the wave operators $\Omega_\pm^U$ for any pair $(H_\alpha^U,-\Delta)$. For simplicity, we shall treat in details only the operator $\Omega_-^U$.

Similarly to \eqref{formuleRui}, the wave operators are expressed in terms of the generalized eigenfunctions $\Psi_\alpha^U$ of $H_\alpha^U$ through the following formula, for $f\in \H$, $r\in \R_+$ and $\theta\in [0,2\pi)$:
\begin{equation*}
[\Omega_-^U f](r,\theta) := \li\;\; \hbox{$\frac{1}{2\pi}$}\int_{\R_+}\int_0^{2\pi} \kappa\;\!
\Psi^U_\alpha (r,\theta,\kappa,\omega)\;\![\FF f](\kappa,\omega)\;\! \d \omega \;\!\d \kappa\ .
\end{equation*}
Furthermore, the functions $\Psi_\alpha^U$ have been calculated explicitly in \cite{AT}. But before writing the rather complicated formulae obtained in this reference, let us introduce a new decomposition of $\H$.

We set $\H_\2:=\H_0\oplus\H_{-1}$ which is clearly isomorphic to $\G:=\H_r\otimes \C^2$, and consider the decomposition $\H =\H_\2 \oplus \H_\2^\bot$. It easily follows from \cite{AT} that for any $U$, the operator $\Omega_-^U$ is reduced by this decomposition, and that the restriction of the wave operator $\Omega_-^U$ to $\H_\2^\bot$ is equal to $\Omega_-^{\AB}$. More generally, this is a consequence to the fact that the functions $\psi_\pm^0$ and $\psi_\pm^{-1}$ introduced in Section \ref{secex} belong to $\H_\2$. Since $\Omega_-^{\AB}$ has already been analyzed in the previous section, we shall concentrate only on the restriction of $\Omega_-^U$ to $\H_\2$.

For that purpose, let us recall the explicit form of $\Psi_\alpha^U$ restricted to $\H_\2$. It is proved in \cite{AT} that, modulo our rearrangement, one has:
\begin{eqnarray*}
\hbox{$\frac{1}{2\pi}$} \;\! \Psi_\alpha^U(r,\theta,\kappa,\omega)\Big|_{\H_\2} &=& \sum_{m \in\{0,-1\}}i^{|m|}\;\!e^{i\delta_m^\alpha}\;\! J_{|m+\alpha|}(\kappa\;\!r)\;\!\phi_m(\theta)\;\! \overline{\phi_m(\omega)} \\
&& + \Big[\hbox{$\frac{1}{2}$}\;\!i^\alpha\;\!H_\alpha^{(1)}(\kappa r)\Big]\;\!4\;\!i\;\!\cos \big(\hbox{$\frac{\pi}{2}$}\alpha \big)\;\!p_{00}(\kappa) \;\!(-\kappa^2)^\alpha\;\!\phi_0(\theta) \;\!\overline{\phi_0(\omega)} \\
&& + \Big[\hbox{$\frac{1}{2}$}\;\!i^\alpha\;\!H_\alpha^{(1)}(\kappa r)\Big]\;\!2\;\!e^{2i\pi \alpha}\;\!\sqrt{2\sin (\pi \alpha)}\;\!p_{-10}(\kappa)\;\!\kappa \;\!\phi_0(\theta) \;\!\overline{\phi_{-1}(\omega)} \\
&& -\Big[\hbox{$\frac{1}{2}$}\;\!i^{1-\alpha}\;\!H_{1-\alpha}^{(1)}(\kappa r)\Big]\;\!2\;\!e^{- 2i\pi\alpha}
\;\!\sqrt{2\sin (\pi \alpha)}\;\!p_{0-1}(\kappa)\;\!\kappa \;\!\phi_{-1}(\theta) \;\!\overline{\phi_0(\omega)}\\
&& -\Big[\hbox{$\frac{1}{2}$}\;\!i^{1-\alpha}\;\!H_{1-\alpha}^{(1)}(\kappa r)\Big]\;\!4\;\!i\;\!\sin \big(\hbox{$\frac{\pi}{2}$}\alpha \big)\;\!p_{-1-1}(\kappa) \;\!(-\kappa^2)^{1-\alpha}\;\!\phi_{-1}(\theta) \;\!\overline{\phi_{-1}(\omega)}\ ,
\end{eqnarray*}
where $p_{jk}$ are functions explicitly calculated in \cite{AT}, and $H_\nu^{(1)}$ is the Hankel function of the first kind and of order $\nu$. We mention that the functions $p_{jk}$ depend on $\alpha$ and $U$.
In order to rewrite this expression and the wave operator $\Omega_-^U$ in a more friendly form, let us introduce a matrical notation: We set for $\kappa$ and $r$ in $\R_+$:
\begin{equation*}
\Tm_\alpha(\kappa\;\!r)=\left(\begin{array}{cc}
\hbox{$\frac{1}{2}$}\;\!i^\alpha\;\!H_\alpha^{(1)}(\kappa r) & 0 \\
0 & \hbox{$\frac{1}{2}$}\;\!i^{1-\alpha}\;\!H_{1-\alpha}^{(1)}(\kappa r)
\end{array}\right)\ ,
\end{equation*}
and
\begin{equation*}
\Sm_\alpha^U(\kappa)=\left(\begin{array}{cc}
4 i\cos\big(\hbox{$\frac{\pi}{2}$}\alpha\big)\;\!p_{00}(\kappa)
\;\!(-\kappa^2)^\alpha
& 2\;\!e^{2i\pi\alpha}\;\!\sqrt{2\sin(\pi\alpha)}\;\!p_{-10}(\kappa) \;\!\kappa \\
-2\;\!e^{-2i\pi\alpha}\;\!\sqrt{2\sin(\pi\alpha)}\;\!p_{0-1}(\kappa) \kappa &
-4 i\sin\big(\hbox{$\frac{\pi}{2}$}\alpha\big)\;\!p_{-1-1}(\kappa) \;\!(-\kappa^2)^{1-\alpha}
\end{array}\right)\ .
\end{equation*}
By using this notation and the isomorphism between $\H_\2$ and $\G$, the restriction of wave operator to $\H_\2$, seen as a map from $\G$ to $\G$, can be rewritten for $f\equiv \big(\begin{smallmatrix} f_0 \\ f_{-1} \end{smallmatrix} \big)\in \G$ and $r \in \R_+$ as:
\begin{equation}\label{labellef}
[\Omega_-^U f](r) = [\Omega_-^{\AB}f](r)+\li\int_{\R_+}\kappa\;
\Tm_\alpha(\kappa\;\!r)\;\!\Sm_\alpha^U(\kappa)\;\![\FF f](\kappa) \;\!\d \kappa\ ,
\end{equation}
where $\FF f =\Big(\begin{smallmatrix} \FF_0 f_0 \\ \FF_{-1} f_{-1} \end{smallmatrix} \Big)$.

In the remaining part of this section, we shall show that the second term can be rewritten as a product of a function of $A$ and a function of $-\Delta$.

\subsection{The operator $T_m$}

We consider first the function of the dilation group. The construction is very similar to the one already encountered in Section \ref{secABop} for the original Aharonov-Bohm operator. For that purpose, let us consider
for $m \in \{0,-1\}$, $g \in C^\infty_c(\R_+)$ and $r \in \R_+$ the following equalities:
\begin{eqnarray}\label{eqsec}
\nonumber [T_mg](r)&:=& s-\lim_{N\to \infty}\hbox{$\frac{1}{2}$}\;\!i^{|m+\alpha|}\int_{1/N}^N \kappa\;\!H^{(1)}_{|m+\alpha|}(\kappa\;\!r)\;\![\FF_m g](\kappa)\;\!\d \kappa\\
\nonumber &=&  s-\lim_{N\to \infty}\hbox{$\frac{1}{2}$} \;\!e^{-i\delta_m^\alpha}\int_{1/N}^N  \kappa \;\!H^{(1)}_{|m+\alpha|}(\kappa\;\!r)
\Big[\int_0^\infty s\;\! J_{|m|}(s\;\!\kappa)\;\!g(s)\;\!\d s\Big]\d \kappa \\
\nonumber &=& s-\lim_{N\to \infty}\hbox{$\frac{1}{2}$}\;\!e^{- i\delta_m^\alpha}\int_0^\infty s\;\!
\Big[\int_{1/N}^N \kappa\;\! H^{(1)}_{|m+\alpha|}(\kappa\;\!r) \;\!J_{|m|}(s\;\!\kappa)\;\!\d \kappa\Big]g(s)\;\! \d s \\
\nonumber &=& s-\lim_{N\to \infty}\hbox{$\frac{1}{2}$}\;\!e^{- i\delta_m^\alpha}\int_0^\infty \hbox{$\frac{s}{r}$}\;\!
\Big[\int_{r/N}^{Nr} \kappa\;\! H^{(1)}_{|m+\alpha|}(\kappa) \;\!J_{|m|}(\hbox{$\frac{s}{r}$}\;\!\kappa)\;\!\d \kappa\Big]g(s)\;\! \hbox{$\frac{\d s}{r}$} \\
&=& \hbox{$\frac{1}{2}$}\;\!e^{- i\delta_m^\alpha} \int_0^\infty \hbox{$\frac{s}{r}$}\;\!\Big[\int_{0}^\infty \kappa\;\! H^{(1)}_{|m+\alpha|}(\kappa) \;\!J_{|m|}(\hbox{$\frac{s}{r}$}\;\!\kappa)\;\!\d \kappa\Big]g(s)\;\! \hbox{$\frac{\d s}{r}$}
\end{eqnarray}
where the last term has to be understood in the sense of distributions on $\R_+$.

As in the previous section, by comparing \eqref{eqsec} with \eqref{formuleJen}, one observes that this operator is equal, at least on a dense set in $\H_r$, to $\tilde{\varphi}_m(A)$ for a function $\tilde{\varphi}_m$ whose inverse Fourier transform satisfies for $y\in\R$:
\begin{eqnarray*}
\check{\tilde{\varphi}}_m (y) &=&\hbox{$\frac{1}{2}$}\;\!
\sqrt{2\pi}\;\!e^{- i\delta_m^\alpha}\;\!e^{-y}\;\!\Big[
\int_0^\infty \kappa\;\! H^{(1)}_{|m+\alpha|}(\kappa) \;\!J_{|m|}(e^{-y}\;\!\kappa)\;\!\d \kappa\Big] \\
&=&\hbox{$\frac{1}{2}$}\;\!
\sqrt{2\pi}\;\!e^{- i\delta_m^\alpha}\;\!e^{y}\;\!\Big[
\int_0^\infty \kappa\;\! H^{(1)}_{|m+\alpha|}(e^y\;\!\kappa) \;\!J_{|m|}(\kappa)\;\!\d \kappa\Big]\ .
\end{eqnarray*}
And again, the distribution between brackets has been explicitly computed in \cite[Prop.~1]{KR_WS}. We recall first the general result.

\begin{Proposition}\label{surImunu}
For any $\mu,\nu\in \R$ satisfying $\nu + 2 >|\mu|$ and $s \in \R_+$ one has
\begin{eqnarray}\label{horreur1}
& \int_0^\infty \kappa \;\!H^{(1)}_\mu (s\kappa) \;\!J_\nu(\kappa)\;\! \d \kappa\ = e^{i\pi(\nu-\mu)/2}\;\!\delta(s-1)  +  \hbox{$\frac{2}{i\pi}$} \;\!e^{i\pi(\nu-\mu)/2}\;\! s^{-1} \;\!\Pv\;\!\big(\hbox{$\frac{1}{\frac{1}{s}-s}$}\big) &\\
\nonumber & + \hbox{$\frac{2}{i\pi}$} \;\!e^{i\pi(\nu-\mu)/2}\;\!\big(\hbox{$\frac{s^{-1}}{\frac{1}{s}-s}$}\big)\;\!
\;\!\Big[s^{-\nu}\hbox{$\frac{\Gamma(\frac{\nu+\mu}{2}+1)\;\!\Gamma(\frac{\nu-\mu}{2}
+1)}{\Gamma(\nu+1)}$}\;\!
\F21\big(\hbox{$\frac{\nu+\mu}{2}$},\hbox{$\frac{\nu-\mu}{2}$}
;\nu+1;s^{-2}\big)-1 \Big]&
\end{eqnarray}
as an equality between two distributions on $\R_+$. The last term belongs to $L^1_{\rm loc}(\R_+)$.
\end{Proposition}

We now state the main properties of the operator $T_m$:

\begin{Proposition}
For $m \in \{0,-1\}$, one has
$T_m = \tilde{\varphi}_m(A)$
with $\tilde{\varphi}_m \in C\big([-\infty,+\infty],\C\big)$. Furthermore these functions satisfy $\tilde{\varphi}_m(-\infty)=0$ and $\tilde{\varphi}_m(+\infty)=1$.
\end{Proposition}

\begin{proof}
Let us define the following distributions for $y \in \R$:
\begin{equation*}
\check{\tilde{\varphi}}_{m,1}(y) = \hbox{$\frac{1}{2}$}\;\!\sqrt{2\pi}\;\!\delta(y), \qquad
\check{\tilde{\varphi}}_{m,2}(y) = \hbox{$\frac{1}{2}$}\;\!i\;\!\sqrt{\hbox{$\frac{2}{\pi}$}}\;\! \Pv\;\!\big(\hbox{$\frac{1}{\sinh(y)}$}\big)
\end{equation*}
and
\begin{equation*}
\check{\tilde{\varphi}}_{m,3}^\pm(y) =\hbox{$\frac{1}{2}$}\;\!i\;\!\sqrt{\hbox{$\frac{2}{\pi}$}}\;\!
\hbox{$\frac{1}{\sinh(y)}$}\;\!\Big[
e^{-\nu y}\hbox{$\frac{\Gamma(\frac{\nu+\mu}{2}+1)\;\!\Gamma
(\frac{\nu-\mu}{2}+1)}{\Gamma(\nu+1)}$}\;\!
\F21\big(\hbox{$\frac{\nu+\mu}{2}$},\hbox{$\frac{\nu-\mu}{2}$}
;\nu+1;e^{-2y}\big)-1\Big]\ ,
\end{equation*}
where the notation $\mu = |m+\alpha|$ and $\nu = |m|$ has been used for shortness. The sum of these distributions is clearly equal to $\check{\tilde{\varphi}}_m$, and it is well know that $\tilde{\varphi}_{m,1}+ \tilde{\varphi}_{m,2} = \hbox{$\frac{1}{2}$}\;\! \big[1 + \tanh\big(\hbox{$\frac{\pi}{2}\cdot$}\big)\big]$. One can already observe that these terms give the correct values at $\pm \infty$.

For $\tilde{\varphi}_{m,3}$, it follows from Proposition \ref{surImunu} that $\check{\tilde{\varphi}}_{m,3}$ belongs $L^1_{\rm loc}(\R)$. For $y \in [0,+\infty)$, the hypergeometric function is bounded, and therefore the map $y \mapsto \check{\tilde{\varphi}}_{m,3}(y)$ has an exponential decrease as $y \to + \infty$, driven by the inverse of the hyperbolic sinus.
For $y \to -\infty$, an asymptotic development of the hypergeometric function is necessary. Borrowing such a development from \cite[Sec.~15.3]{AS}, one easily obtains that the leading term of $\check{\tilde{\varphi}}_{m,3}(y)$ for $y \to -\infty$ is of the form $e^{-y(|m+\alpha|-1)}$, which is exponentially decreasing if and only if $m \in \{0,-1\}$. It thus follows that $\check{\tilde{\varphi}}_{m,3}$ belongs to $L^1(\R)$, and its Fourier transform is then in $C_0(\R)$. The statement follows then from the density of $C^\infty_c(\R_+)$ in $\H_r$.
\end{proof}

\subsection{New formula for the wave operators}

We shall now collect all information obtained so far, and propose a new formula for $\Omega_-^U$.

Since the wave operators $\Omega_\pm^U$ are reduced by the decomposition of $\H$ into $\H_\2 \oplus \H_\2^\bot$, so does the scattering operator $S_\alpha^U \equiv S^U_\alpha(-\Delta):=\big(\Omega_+^U\big)^* \Omega_-^U$. Furthermore, by looking at the restriction of $S^U_\alpha$ to $\H_\2$ and by considering it as a map from $\G$ to $\G$, one naturally observes that there exists a close relation between this map and the family $\Sm_\alpha^U(\cdot)$ introduced before. Indeed, by comparing the expression of $\Sm_\alpha^U(\cdot)$ with the formula obtained in \cite{AT} for the scattering amplitude $f^U_\alpha$, and by taking into account the relation between the scattering amplitude and the scattering operator \cite{T1,Rui}, one observes that the following equality holds on $\G$:
\begin{equation*}
\Sm_\alpha^U(\sqrt{-\Delta}) = S_\alpha^U(-\Delta) -
\Big( \begin{smallmatrix}e^{-i\pi\alpha} & 0 \\ 0 & e^{i\pi\alpha} \end{smallmatrix}\Big)
\end{equation*}
where $\Sm_\alpha^U(\sqrt{-\Delta})$ is given by $\FF^* \Sm_\alpha^U(\k) \FF$ and $\Sm_\alpha^U(\k)$ is the operator of multiplication by $\Sm_\alpha^U(\cdot)$ in $\G$.

The following new description of the wave operators is now an easy consequence of the above observation and of the results obtained before for $\varphi_m$ and $\tilde{\varphi}_m$.

\begin{Theorem}\label{thm2}
For any $U$, the restriction of the wave operator $\Omega_-^U$ to $\H_\2$, seen as a map from $\G$ to $\G$, satisfies the equality
\begin{equation}\label{yoyo}
\Omega_-^U = \Big(
\begin{smallmatrix}\varphi^-_0(A) & 0 \\ 0 & \varphi^-_{-1}(A) \end{smallmatrix}\Big) +
\Big(
\begin{smallmatrix}\tilde{\varphi}_0(A) & 0 \\ 0 & \tilde{\varphi}_{-1}(A) \end{smallmatrix}\Big)
\Big[S_\alpha^U(-\Delta) -
\Big( \begin{smallmatrix}e^{-i\pi\alpha} & 0 \\ 0 & e^{i\pi\alpha} \end{smallmatrix}\Big)\Big]\ .
\end{equation}
\end{Theorem}

\begin{proof}
It has been proved in Section \ref{secABop} that the term $\Omega_-^{\AB}$ in \eqref{labellef} takes the form of the first term on the r.h.s. of \eqref{yoyo}. Then, the second term of \eqref{labellef} is also equal to
\begin{eqnarray*}
&&\li\int_{\R_+}\kappa\;
\Tm_\alpha(\kappa\;\!r)\Big[\FF\;\!\big(\FF^*\;\!\Sm_\alpha^U(\k)\;\!\FF\big) f\Big](\kappa) \;\!\d \kappa\ \\
=&& \Big(
\begin{smallmatrix}T_0 & 0 \\ 0 & T_{-1} \end{smallmatrix}\Big)
\Big[\big(\FF^*\;\!\Sm_\alpha^U(\k)\;\!\FF\big)f\Big](r)\ ,
\end{eqnarray*}
which implies the statement.
\end{proof}

\begin{Remark}
A similar formula holds for $\Omega_+^U$. The precise formula can either be calculated again from $\Psi_\alpha^U$ or from the equality $\Omega_+^U = \Omega_-^U\;\!\big(S_\alpha^U(-\Delta)\big)^*$.
\end{Remark}

\end{document}